\documentclass[sigconf]{acmart}
\acmConference[ICSE 2022]{The 44th International Conference on Software Engineering}{May 21–29, 2022}{Pittsburgh, PA, USA}

\usepackage{listings}
\usepackage{graphicx}
\usepackage{multirow}
\usepackage{wrapfig,booktabs}
\usepackage{subcaption}
\definecolor{javared}{rgb}{0.6,0,0} 
\definecolor{javagreen}{rgb}{0.25,0.5,0.35} 
\definecolor{javapurple}{rgb}{0.5,0,0.35} 
\definecolor{javadocblue}{rgb}{0.25,0.35,0.75} 
\lstdefinelanguage{JavaScriptColor}{
  keywords={break, case, catch, const, continue, debugger, default, delete, do, else,
    finally, for, function, if, in, instanceof, new, of, return, switch, this,
    throw, try, typeof, var, void, while, with,
    class, interface, implements, public, private, constructor
  },
  morecomment=[l]{//},
  morecomment=[s]{/*}{*/},
  morestring=[b]',
  morestring=[b]",
  keywordstyle=\color{javapurple}\bfseries,
  identifierstyle=\color{black},
  commentstyle=\color{javagreen}\ttfamily,
  numberstyle=\color{javared}\ttfamily,
  stringstyle=\color{javared}\ttfamily,
  sensitive=false,
  numbers=left,
  stepnumber=1,
  escapeinside={/*\#}{\#*/},
}
\newcommand{\filename}[1]{\texttt{#1}}
\newcommand{\pkgspec}[1]{\texttt{#1}}
\newcommand{\code}[1]{\texttt{#1}}
\newcommand{\amalfi}{\textsc{Amalfi}}
\newcommand{\barcolor}{\text{red}}

\begin{document}
\title{Practical Automated Detection of Malicious npm Packages}

\author{Adriana Sejfia}
\email{sejfia@usc.edu}
\affiliation{%
  \institution{University of Southern California}
  \city{Los Angeles}
  \country{USA}
}

\author{Max Sch\"afer}
\email{max-schaefer@github.com}
\affiliation{%
  \institution{GitHub}
  \city{Oxford}
  \country{UK}
}

\begin{abstract}
    The npm registry is one of the pillars of the JavaScript and TypeScript
ecosystems, hosting over 1.7 million packages ranging from simple utility
libraries to complex frameworks and entire applications. Each day, developers
publish tens of thousands of updates as well as hundreds of new packages.
Due to the overwhelming popularity of npm, it has become a prime target for
malicious actors, who publish new packages or compromise existing packages to
introduce malware that tampers with or exfiltrates sensitive data from users who
install either these packages or any package that (transitively) depends on
them. Defending against such attacks is essential to maintaining the integrity
of the software supply chain, but the sheer volume of package updates makes
comprehensive manual review infeasible.
We present \amalfi, a machine-learning based approach for automatically detecting
potentially malicious packages comprised of three complementary techniques. We
start with classifiers trained on known examples of malicious and benign
packages. If a package is flagged as malicious by a classifier, we then check
whether it includes metadata about its source repository, and if so whether the
package can be reproduced from its source code. Packages that are reproducible
from source are not usually malicious, so this step allows us to weed out false
positives. Finally, we also employ a simple textual clone-detection technique to
identify copies of malicious packages that may have been missed by the
classifiers, reducing the number of false negatives.
\amalfi\ improves on the state of the art in that it is lightweight,
requiring only a few seconds per package to extract features and run the
classifiers, and gives good results in practice: running it on 96287 package
versions published over the course of one week, we were able to identify 95
previously unknown malware samples, with a manageable number of false positives.

\end{abstract}

  
\ccsdesc[500]{Security and privacy~Malware and its mitigation}

\keywords{supply chain security, malware detection}

\newcommand{\maxsch}[1]{{\color{blue}(Max: #1)}}
\newcommand{\adriana}[1]{{\color{teal}(Adriana: #1)}}

\maketitle

\section{Introduction}\label{sec:intro}
npm\footnote{\url{https://npmjs.com}} is a system for publishing and consuming
software packages for JavaScript and TypeScript. While initially closely
associated with the Node.js platform\footnote{\url{https://nodejs.org}} and
back-end JavaScript applications, it is not architecturally tied to Node.js, and
has also found widespread use with web applications and on other platforms.

The core concept of npm is the \emph{package registry}, which is a database of
JavaScript packages with associated metadata. While some organizations and
enterprises host their own registries, by far the best-known registry is the
public npm registry, accessible via the npm website, which also provides
facilities for browsing and searching for packages, as well as viewing their
metadata. In this paper, we exclusively concern ourselves with the public
registry.

As of early September 2021, npm's package registry hosts over 1.7 million packages. Some of these
are private packages that are only accessible to specific users or
organizations, but most of them are public, and it is these public packages that
are our focus. Over the course of a single week, developers publish around
100,000 public package versions, including both new packages and updated
versions of existing packages. Historic versions of a package remain available
on the registry unless they are explicitly removed either by the package
maintainer or by npm staff, allowing dependent packages to rely on specific
older versions of a package, for example to make use of an API that has been
removed in the latest version.

Most developers interact with the registry through a command-line interface such
as the npm CLI\footnote{\url{https://www.npmjs.com/package/npm}} or
yarn,\footnote{\url{https://www.npmjs.com/package/yarn}} which can be used to
download a particular version of an existing package and install it locally, or
to publish a new package or a new version of an existing package to the
registry. When installing a package version, the package manager will first
recursively install the dependencies of that package (unless they are already
installed); download the tarball containing the package from the registry;
unpack the tarball in the installation directory; and finally run any
installation scripts specified by the package. These scripts are free-form shell
scripts that typically perform setup tasks such as downloading additional
artifacts not bundled with the package itself. Due to the transitive nature of
package installation, popular packages are downloaded very frequently; for
example, the \texttt{chalk}
package\footnote{https://www.npmjs.com/package/chalk} (which offers support for
coloring terminal output) was downloaded almost 89 million times a week at the
time of writing.

Publishing a new package or package version is the dual to this process: anyone
authenticated through the npm website can create a new package, thereby becoming
its maintainer, and maintainers can publish new versions at any time by simply
uploading a tarball to the registry. While packages can indicate a repository
hosting their source code, this information is optional. It is the maintainer's
responsibility to run any necessary build steps (such as compiling TypeScript to
JavaScript, bundling and minifying, etc.) before publishing; the registry simply
hosts the tarball and is largely agnostic to its content.

The overwhelming popularity of npm and the central role it plays in the software
supply chain for JavaScript and TypeScript (which, in turn, are among the most
widely-used programming languages at present) has long made it a favorite target
for attackers attempting to publish malicious package versions that tamper with
or exfiltrate data from the machines they are installed on, perform parasitical
computations such as Bitcoin mining, or other malicious activities.

\begin{sloppypar}
Recent examples include high-profile incidents such as the
\pkgspec{eslint-scope}
compromise,\footnote{\url{https://eslint.org/blog/2018/07/postmortem-for-malicious-package-publishes}}
where attackers managed to steal the credentials of a maintainer of a popular
package, allowing them to publish a new malicious version of the package that
uploaded user credentials to a server upon installation; the \pkgspec{event-stream}
backdoor,\footnote{\url{https://snyk.io/blog/a-post-mortem-of-the-malicious-event-stream-backdoor}}
where social-engineering techniques were used to gain maintainer status and then
launch a similar attack; and a steady stream of smaller incidents since then.
\end{sloppypar}

\begin{sloppypar}
The malicious package versions were quickly removed by npm staff from the
registry upon detection, but, in the case of \pkgspec{eslint-scope} and
\pkgspec{event-stream}, not before being installed several million times. While
the number of affected users is much smaller in most cases, the frequency with
which such incidents occur still poses a significant danger to the software
supply chain, both in terms of concrete damage to its users, and in terms of
reputational damage that could potentially impede the flourishing not just of
npm but also the wider open-source ecosystem.
\end{sloppypar}

Addressing this problem at a fundamental level would arguably require
significant changes to npm, perhaps including a more secure package-publishing
model to prevent malicious packages from reaching the registry in the first
place, and/or the Node.js platform, perhaps with access-control enforcement to
limit the damage a malicious package can do when it is installed.

Our aim in this paper is more modest: without making any changes to the
fundamentals of npm, we want to detect potentially malicious package versions as
quickly as possible, and report them to a human auditor for take-down.

To be practically useful, then, our approach has to satisfy at least three
requirements: it has to be \emph{automated}, since the sheer number of packages
renders manual audits infeasible; \emph{efficient} to keep up with the speed at
which new versions are published; and \emph{accurate} to avoid flagging benign
packages or missing malicious ones.

We achieve this by combining three complementary techniques into one system,
which we call \amalfi:\footnote{Short for ``''\textbf{A}utomated \textbf{mal}icious package \textbf{fi}nder''.}
\begin{enumerate}
    \item machine-learning \emph{classifiers} trained on labelled examples of
    malicious and benign packages, utilizing features that record changes in
    the APIs the package uses as well as package metadata extracted using a
    lightweight syntactic scan;
    \item a \emph{reproducer} that rebuilds a package from source and compares
    the result with the version published in the registry;
    \item a \emph{clone detector} that finds (near-)verbatim copies of known
    malicious packages.
\end{enumerate}

Our feature selection, discussed in more detail below, is motivated by the
observation (borrowed from Garrett et~al.~\cite{garrett19}) that malicious
packages tend to make use of distinctive capabilities of the JavaScript language
(such as runtime code generation), the underlying platform (such as access to
the file system or the network), and the npm package manager (such as install
scripts). While none of these features are dead giveaways by themselves, in
combination they are worthy of closer inspection, especially if a package
suddenly starts using capabilities it has never used before. For example, the
above-mentioned \pkgspec{eslint-scope} package uses runtime code generation and
an install script in its (malicious) version 3.7.2, capabilities it had never
used before.

By training on a corpus of malicious and benign packages provided to us by npm,
our classifiers learn to distinguish typical (and therefore most likely
harmless) feature changes from atypical (and therefore suspicious) ones. The
choice of classifiers is constrained by the small size of the corpus, which
contains fewer than 2000 samples; we experimented with three different
techniques: decision trees, Naive Bayesian classifiers, and one-class SVMs.

To eliminate false positives, we borrow another insight from the
literature~\cite{vu20, vu21, goswami20}: malicious package versions tend not to
have their source code publicly available, in order to avoid
detection.\footnote{In fact, we are not aware of a single counterexample to this
rule.} Consequently, being able to reproduce a package version from its source
code is a good indicator that it is benign. As has been noted
previously~\cite{goswami20}, even perfectly benign packages may fail to
reproduce for a variety of reasons, but this is acceptable in our case since we
are only using this criterion to filter out benign packages erroneously flagged
as malicious, not to detect new ones.

Finally, we note that attackers often publish multiple textually identical
copies of one and the same malicious package under different names. However,
since package metadata may be different, our classifiers sometimes fail to spot
these copies. We use a simple clone detector that hashes the contents of a
package (minus the package name and version, which are always unique) to
eliminate this source of false negatives.

\begin{figure}

\centering
\includegraphics[width=9cm]{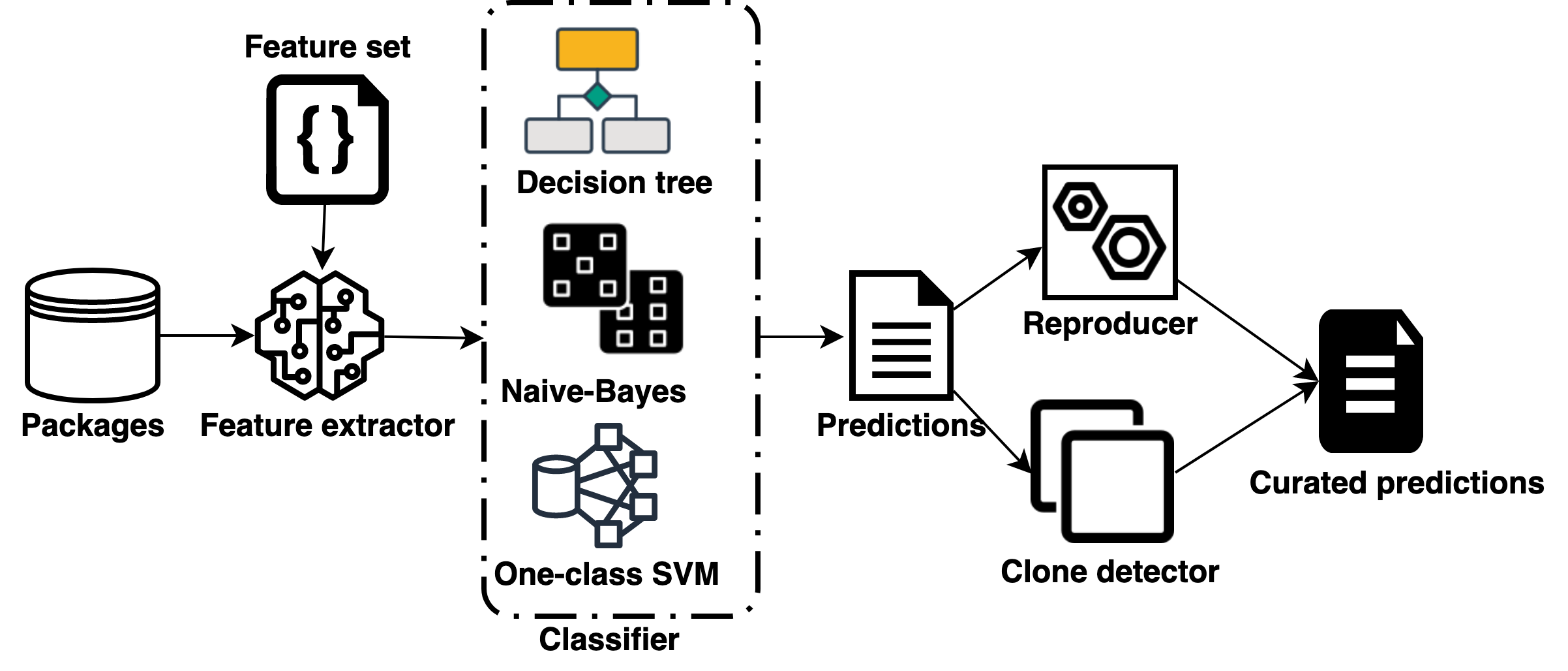}

\caption{Overview of \amalfi}
\Description[Overview of the individual components of \amalfi]{npm packages are represented as feature vectors,
which are then presented to three types of classifiers (Decision tree, Naive Bayes, and one-class SVM) to identify
potentially malicious packages. These predictions are then post-processed by the reproducer to filter out false positives
and by the clone detector to detect additional true positives.}
\label{fig:approach}
\end{figure}

An overview of how these different components work together can be seen in
Figure~\ref{fig:approach}.

To motivate our approach more carefully, we show two typical examples of
malicious packages in Section~\ref{sec:background} and discuss how we detect
them. Section~\ref{sec:approach} provides details on feature selection and
extraction for the classifier, as well as an overview of the reproducer and
clone detector. We evaluate \amalfi\ in Section~\ref{sec:evaluation} in a
large-scale experiment on newly published npm packages to demonstrate its
ability to find previously unknown malicious packages, and in a cross-validation
experiment to evaluate precision and recall. We discuss the
results in Section~\ref{sec:discussion}, survey related work in
Section~\ref{sec:related}, and outline conclusions and future directions in
Section~\ref{sec:conclusion}.

In summary, the significance of our contributions is as follows:
\begin{itemize}
    \item We present \amalfi, an automated approach for detecting malicious npm packages
    that uses a novel combination of techniques and shows solid results in
    practice, identifying 95 previously unknown malicious packages. Among the
    three different classifiers we consider, the decision tree performs best,
    though the others also contribute findings.
    \item \amalfi\ is efficient, usually taking only a few seconds per
    package to extract features and run the classifier. We also show
    that retraining the classifiers is cheap, thus allowing continuous
    improvements to be made as more and more results are triaged.
    \item The false-positive rate, while initially quite high, drops
    significantly as the classifiers are retrained on more data, with fewer than
    one in a thousand packages being flagged spuriously. A cross-validation
    experiment on our training set also shows that the decision tree achieves
    over 40\% recall, suggesting that its false-negative rate is reasonable.
    Supplementary materials including experimental data and results are publicly
    available at \url{https://doi.org/10.5281/zenodo.5908852} and
    \url{https://github.com/githubnext/amalfi-artifact}.
\end{itemize}

While the building blocks we use have been proposed before, the novelty of our
approach lies in their combination, and a more thorough exploration and
evaluation of the design space.

\section{Background}\label{sec:background}
To set the scene, we discuss two representative examples of real-world malicious
package versions that were (manually) detected and removed from the registry,
and then explain how we could have identified them automatically.

For the purposes of this paper, we define a \emph{malicious package version} to
be a specific version of an npm package that contains code that implements
malicious behavior including (but not limited to) exfiltrating sensitive or
personal data, tampering with or destroying data, or performing long-running or
expensive computations that are not explicitly documented. In particular, we
consider a package version to be malicious even if the malicious code it
contains is disabled or broken. Moreover, in line with npm's Acceptable Content
Policy\footnote{\url{https://docs.npmjs.com/policies/open-source-terms\#acceptable-content}}
we include in our definition malicious behavior that is ostensibly done for
research purposes. For brevity, we will often use the term ``malicious
package'', the ``version'' part being understood.

From an attacker's perspective, there are three steps to delivering malware
through npm: (1) publish a malicious package version; (2) get users to install
it; and (3) get them to run the malicious code.

The easiest way to go about (1) is to publish a completely new package. A
classic way of achieving (2) in this scenario is
\emph{typosquatting}~\cite{tschacher16} whereby the name chosen for the new
package is very similar to the name of a popular existing package; a user who
accidentally misspells the name of the popular package will then end up
inadvertently installing the malicious package instead. A more sophisticated
approach is \emph{dependency confusion}~\cite{birsan21}: the attacker identifies
dependencies on a package hosted in a private registry, and then publishes a
malicious package with the same name and a higher version number on the public
npm registry; clients of the private package may then end up installing the
malicious package instead. Finally, there have been cases of attackers
publishing an initially benign and useful package, getting it added as a
dependency to a popular target package, and then publishing a malicious
version~\cite{baldwin19}.

An alternative, more laborious strategy to achieve (1) is for the attacker to
compromise an existing popular package by gaining maintainer access (for example
by stealing maintainer credentials or by social engineering as described in
Section~\ref{sec:intro}), and then publishing a new, malicious version of that
package. In this case, (2) is easy since the package already has many users who
will (either explicitly or implicitly) upgrade to the malicious version.

Finally, a common tactic to achieve (3) in either scenario is to use
installation scripts which (as explained above) are run during installation and
can execute arbitrary code. However, the commands run by installation scripts
are by default logged to the console, increasing the risk of detection. Hence a
more careful attacker may instead choose to hide their malicious code in some
frequently executed bit of functionality in the main body of the package.

A typical example of a package employing typosquatting is \pkgspec{mogodb}, a
putative typo for the highly popular \pkgspec{mongodb} package, which is
currently seeing around two million installations per week. Two versions of
\pkgspec{mogodb}, numbered 3.1.8 and 3.1.9, were published within less than
a millisecond of each other on 1 August 2019, and identified as malicious
and taken down a few minutes later.

As shown in Figure~\ref{fig:mogodb}~(a), the \filename{package.json} manifest
file of the package registers a \code{postinstall} script to be run after
package installation, which executes the \filename{test.js} script included in
the package. That script, shown in Figure~\ref{fig:mogodb}~(b), harvests the
hostname of the machine on which the package was installed, and sends it off to
a remote host controlled by the attacker. While the information being stolen in
this case is not highly sensitive, this is clearly malicious behavior.

\begin{figure}
\begin{lstlisting}
{
  "name": "mogodb",
  "version": "3.1.9",
  "scripts": {
    "postinstall": "node test.js",
    ...
  },
  ...
}
\end{lstlisting}
(a) \filename{package.json}
\begin{lstlisting}
var remote = "https://attacker.controlled.host/";
var host = require("os").hostname();
require("request")(remote + "?h=" + host, function() {});
\end{lstlisting}
(b) \filename{test.js}
\caption{Malicious code in \pkgspec{mogodb@3.1.9} (simplified)}\label{fig:mogodb}
\end{figure}

Even before studying the package implementation in detail, a human auditor might
notice features of the package that make it seem worthy of closer scrutiny, such
as the presence of a postinstall script, the usage of the packages \pkgspec{os}
and \pkgspec{request}, and most of all the extremely short time span between the
publication of the two versions. While the former two features are not by
themselves suspicious, their combination with each other and with the third
feature strongly suggests a malicious package.

A typical example of a compromised package is \pkgspec{jasmin}, a web framework
that was moderately popular at one point but has not seen active development in
a number of years. Versions 0.0.1 and 0.0.2 of \pkgspec{jasmin} are benign, but
version 0.0.3, presumably published by a malicious actor, contains the code
shown in Figure~\ref{fig:jasmin}, which traverses all forms contained in an HTML
document looking for password fields and overrides their submit handler to
harvest the content of these fields and send them to an attacker-controlled
host.

\begin{figure}
\begin{lstlisting}
var remote = "https://another.attacker.controlled.host/";
for (var form of document.forms) {
  for (var element of form.elements) {
    if (element.type == "password") {
      form.addEventListener('submit', function() {
        var data = [...this.elements].map(function(elt) {
           return elt.name + ":" + elt.value;
        }).join() + "|" + document.cookie;
        var enc = encodeURIComponent(btoa(data));
        this.action = remote + "?data=" + enc;
      });
      break;
    }
  }
}
\end{lstlisting}
\caption{Malicious code inserted into file \filename{component.js} of \pkgspec{jasmin@0.0.3} (simplified)}\label{fig:jasmin}
\end{figure}

In this case, there are no particularly suspicious features of the package that
might draw the attention of a human auditor: dealing with password fields,
encoding data using \code{encodeURIComponent} for transmission, and accessing
HTTP cookies are all relatively innocent capabilities, and are often used
together. What is immediately suspicious, however, is that none of these three
capabilities were used in the previous version of \pkgspec{jasmin}. Moreover,
the upgrade from 0.0.2 to 0.0.3 is a minor version upgrade, where one would not
expect major new features that might require such new capabilities to be
introduced.

These examples and others like them suggest that a machine-learning based
approach might be able to detect malicious packages based on high-level features
like usage of particular APIs, platform capabilities and package metadata, and
in particular how these features change between versions, without the need for
deep source-code analysis.

\section{Our Approach}\label{sec:approach}
Having motivated what kind of features are interesting for automated
classification, we now describe our feature set in more detail, and then explain
how to extract \emph{single-version features} describing one package version as
well as \emph{change features} capturing the difference in features between two
versions. Next, we discuss our choice of classifiers and their training regimen.
Finally, we give some more details about the other two components of our
approach, the reproducer and the clone detector.

\subsection{Feature set}
Based on manual inspection of known examples of malicious packages, we
determined eleven features of interest. Nine of them are single-version features
that can be extracted from the contents of a single package version, while the
other two intrinsically involve two versions of a package.

The single-version features are as follows, where we group related features into
categories and provide examples of each:

\begin{enumerate}
\item Access to personally-identifying information (PII): credit-card numbers,
passwords, and cookies
\item Access to specific system resources
	\begin{enumerate}
	\item File-system access: reading and writing files
    \item Process creation: spawning new processes
    \item Network access: sending or receiving data
    \end{enumerate}
\item Use of specific APIs
    \begin{enumerate}
	\item Cryptographic functionality
	\item Data encoding using \code{encodeURIComponent} etc.
	\item Dynamic code generation using \code{eval}, \code{Function}, etc.
	\end{enumerate}
\item Use of package installation scripts
\item Presence of minified code (to avoid detection) or binary files (such as
binary executables)
\end{enumerate}

The remaining two features concern two versions of a package, and are the time
between publication of the two versions, and the type of update in
semantic-versioning terms (major, minor, patch, build, or pre-release).

The motivation for considering time between updates is that malicious package
versions often exhibit unusual update patterns, such as multiple versions
published in very rapid succession (as seen in the \pkgspec{mogodb} example in
Section~\ref{sec:background}), or a new version being published after years of
inactivity (which might suggest an account takeover). The update type, on the
other hand, can determine whether a change in some other feature is suspicious
or not, as explained above.

These two features, along with the changes in the values of the nine
single-valued features between one package version and the previous version,
constitute our feature set.

In order to accommodate the first version of a package as well, we introduce a
pseudo-update type representing first versions, consider their time between
updates to be zero, and take the values of the single-version features to be the
remaining change features. This enables us to not only detect malicious updates,
where a previously benign package becomes malicious, but also packages that were
malicious from the start.

\subsection{Feature extraction}
To compute the first four categories of single-version features, we parse each
JavaScript and TypeScript file in the package using
Tree-sitter.\footnote{\url{https://tree-sitter.github.io/tree-sitter/}} We then
use Tree-sitter AST queries to look for syntactic constructs corresponding to
the features, such as string literals containing the keyword \code{password} for
PII access; imports of the \code{fs} module for file-system access; and calls to
\code{eval} and \code{Function} for dynamic code generation.

Similarly, to check for the presence of installation scripts we parse the
\filename{package.json} file and look for definitions of \code{preinstall},
\code{install}, and \code{postinstall} properties.

Minified or binary files tend to have higher entropy than plain source code, so
we compute the Shannon entropy of all files contained in the package and use the
average and standard deviation of the entropy across all files as features.

To compute change features, we use the publication timestamps provided by the
\code{npm view time} command to obtain the time between updates in seconds. We
rely on an off-the-shelf semantic-versioning library to determine the update
type and, for each given version, determine the previous version in
chronological order. Finally, we simply subtract the values of the
single-version features across the two consecutive versions.

\subsection{Classifier training}
Our choice of classifiers is dictated by the corpus of labelled training data we
have available. Since malicious packages are taken down by npm immediately upon
discovery, most known examples of malicious packages are no longer available for
inspection. However, npm kindly agreed to make their archive of 643 malicious
package versions detected up to 29 July 2021 available to us for the purposes of
this study. Out of these packages, 63 are malicious versions of otherwise non-malicious packages, i.e., compromised packages. We added to the original dataset the 1147 benign versions of the same packages
published by the same date, yielding a basic corpus of 1790 labelled samples of
malicious and benign package versions. Since the goal of \amalfi\ is to detect malicious packages,
the basic corpus oversamples malicious packages, i.e., it contains more malicious packages
than we would expect from a similarly-sized random sample of npm packages. This is a common strategy in learning-based approaches~\cite{mohammed2020}. 

It is worth emphasizing that while compromised packages have a much bigger
potential impact on the npm ecosystem, they occur so rarely that there simply is
not enough data to make them the sole focus of our study. Anecdotally, however,
compromised and malicious packages use similar techniques to carry out attacks,
meaning that they share features, which enables \amalfi\ to detect both
types of malicious packages.

Since the number of malicious samples is smaller compared to the total number
of package versions in our dataset and on npm, we had to use learning algorithms that handle
imbalanced data well. Further, due to the novelty of the features in our
approach, we sought a learning algorithm that allowed us to analyze the
importance of the features we selected. In the end, the learning algorithms that
satisfied the constraints were decision trees, Naive Bayesian classifiers, and
One-class Support Vector Machines (SVMs). We picked the first one due to its
ability to explain which features impact the final decision, and the two latter
ones because of their versatility when dealing with imbalanced datasets as seen
in anomaly detection work.

To train the classifiers, we use the \code{sklearn} library for Python. For the
decision tree, we use information gain as the split criterion. For the Naive
Bayesian classifiers, we use the Bernoulli variant which can only deal with
Boolean features, so we omit the discrete features (entropy average and standard
deviation as well as update time), and collapse the others to a value of 1 if
the feature is present, and 0 otherwise. For the SVM, we choose a linear
kernel and train only on benign examples, since the task of this classifier is
to detect outlier versions that are noticeably different from the benign ones.
We determined the $\nu$ parameter of the SVM, which approximates the number of
expected outliers, by conducting a leave-one-out experiment on our basic corpus.
The experiment showed that optimal precision and recall are attained for a $\nu$
value of 0.001, meaning that the classifier expects about one in a thousand
package versions to be malicious.%
\footnote{The source code of our classifier-training scripts and the list of
packages in the basic corpus are included in the supplementary materials.}

\subsection{Reproducer and clone detector}
As explained in Section~\ref{sec:intro}, the reproducer takes a given package
version and then attempts to rebuild the package tarball from source. This is a
heuristic process that may fail for a variety of reasons: while packages can
specify the URL of their source repository in their \filename{package.json}
file, this information is optional and many packages do not provide it, or the
repository is not publicly accessible. Package versions can also specify the git
SHA of the commit they were built from, but again this information is optional.
While there are popular conventions for creating branches or tags with names
reflecting the package version they correspond to, many packages do not follow
these conventions, making it impossible to determine the correct commit. The
build commands to run to produce the package from its source are likewise not
prescribed. Finally, many packages neglect to specify the precise version of the
build tools (such as the TypeScript compiler) they rely on, leading to seemingly
random differences between the reproduced package and the original. For all of
these reasons, the success rate of the reproducer is low in practice as we shall
see, but it still serves a useful purpose as an automated false-positive filter.

The third component of \amalfi\ is a simple clone detector that computes an
MD5 hash of the contents of a package tarball and compares it to a list of
hashes of known malicious packages. When computing the hash, we ignore the
package name and version specified in the \filename{package.json} file, since
these are always unique and would cause spurious misses. No other attempt at
fuzzy matching is made, so only verbatim clones are detected.


\section{Evaluation}\label{sec:evaluation}
\newcommand{\tpheader}{\multicolumn{1}{|c}{\textbf{\# TP}}}
\newcommand{\fpheader}{\multicolumn{1}{|c}{\textbf{\# FP}}}

\begin{table*}[ht]
\begin{tabular}{l|r||r|r|r|r|r|r|r}
\textbf{Date} & \textbf{\# Versions} & \multicolumn{2}{c|}{\textbf{Decision Tree}} & \multicolumn{2}{c|}{\textbf{Naive Bayes}} & \multicolumn{2}{c|}{\textbf{SVM}} & \textbf{Clones} \\
& & \tpheader & \fpheader & \tpheader & \fpheader & \tpheader & \fpheader & \tpheader \\ \hline
July 29 & 23,452 & 34+1 & 932-74 & 13+22 & 1453-107 & 20+5 & 102-11 & 0 \\
July 30 & 13,849 & 2+0 & 22-1 & 0+0 & 6-0 & 0+0 & 14-3 & 0 \\
July 31 & 7,042 & 17+0 & 16-1 & 0+0 & 1-0 & 18+9 & 4-0 & 0 \\
August 1 & 6,050 & 1+0 & 6-2 & 1+0 & 3-2 & 0+0 & 13-0 & 0 \\
August 2 & 13,562 & 2+1 & 17-0 & 4+1 & 12-1 & 0+0 & 10-0 & 1 \\
August 3 & 15,269 & 6+0 & 15-2 & 1+0 & 9-1 & 0+0 & 41-3 & 0 \\
August 4 & 17,063 & 16+2 & 9-1 & 1+0 & 9-1 & 1+0 & 10-1 & 17 \\
\end{tabular}\\[1mm]
\caption{Results from Experiment 1; $+n$ denotes TPs contributed by the clone detector, $-n$ FPs eliminated by the reproducer.}
\label{tab:sliding-window}
\end{table*}

While motivating and presenting the details of our approach above, we have
informally argued that its design makes it practically useable and useful. We
will now back up these claims with an experimental study, which aims to answer
the following three research questions:

\begin{description}
\item[RQ1] Does \amalfi\ find malicious packages in practice? 
\item[RQ2] Is it accurate enough to be useful? 
\item[RQ3] Is training and classification fast enough to be useable? 
\end{description}

To answer these questions, we conducted two experiments: one experiment on a
large set of newly published package versions to assess performance, and one
smaller experiment on labelled datasets to assess accuracy. We will describe the
experiments below.

\subsection{Experiment 1: Classifying newly published packages}
This was a large-scale experiment designed to simulate a realistic scenario for
automated malware detection in which we applied \amalfi\ to all new public
package versions published on the public npm registry over the course of a
single, randomly chosen week from 29 July 2021 to 4 August 2021.

On the first day, we trained our three classifiers on the basic corpus and then
used them to classify the set $N_1$ of all new package versions published that
day. Additionally, we ran our clone detector on the same set to find copies of
malicious packages in the basic corpus, acting as a fourth classifier. This
yielded a set $P_1\subseteq N_1$ of package versions flagged by at least one
classifier. We ran the reproducer on this set to automatically weed out some
false positives, and manually inspected the rest. The manual inspection was
initially conducted by both authors, with each author examining roughly one half
of the flagged versions. The package versions that were found to be malicious by
one author were afterwards verified by the other author. Finally, we reported
the verified malicious packages to the npm security team. All of them were
subsequently taken down, meaning that the npm security experts agreed with our
assessment. As such, we are confident that our manual labeling of malicious
packages is highly accurate.
 
The manual inspection resulted in a partitioning of $P_1$ into
two sets $\mathit{TP}_1$ and $\mathit{FP}_1$ of true positives (i.e.,
genuine malicious packages found by the classifiers) and false positives (i.e.,
benign packages falsely flagged as malicious). As a last step, we ran the clone
detector again to find additional copies of packages in $\mathit{TP}_1$ that
were missed by the classifiers, and added them to $\mathit{TP}_1$.

On the second day, we retrained the classifiers on the basic corpus as well as
the set $N_1$ triaged the previous day, adding $\mathit{TP}_1$ to our set of
labelled malicious packages, and everything else (that is, $N_1\setminus
\mathit{TP}_1$) to the set of benign packages. In other words, for the
purposes of this experiment we assumed that any package \emph{not} flagged by
any of the classifiers was benign. This is not true in general, but the enormous
number of new package versions published each day made it infeasible to inspect
them all, and since we expect the number of malicious packages on any given day
to be low it is not an unreasonable approximation to the unknown ground truth.

As on the first day, we then applied the classifiers to the set $N_2$ of
packages published that day, ran the reproducer on the resulting set $P_2$,
manually inspected the rest, and ran the clone detector to mop up anything that
was missed, yielding a set $\mathit{TP}_2$ of newly identified malicious
packages. On the third day, we retrained the classifiers using the basic corpus
as well as both $N_1$ an $N_2$, and so forth for each subsequent day.%
\footnote{The list of packages considered in this experiment and the results
of the classification are included in the supplementary materials.}

The intuition here is that we want to mimic a usage pattern where results from
the classifiers are inspected by a human auditor, and the classifiers are then
retrained with the additional ground truth obtained in this way.

\subsection{Experiment 2: Classifying labelled data}
While the first experiment can provide insight into the performance of our
approach under real-world conditions and in particular its false-positive rate,
it cannot tell us much about false negatives.

Hence we ran a second experiment, measuring the precision and recall of
\amalfi\ on the basic corpus. Its small size prevented us from separating it in
a train-and-test fashion, so instead we performed a 10-fold cross validation
experiment, repeatedly training the classifiers on 90\% of the corpus and
measuring precision and recall on the remaining 10\%. Given the imbalance in our
dataset, we used stratified sampling to maintain the distribution of malicious
and benign versions for each fold. 

Furthermore, we also measured precision and recall on a labelled dataset from
recent work by Duan et al. describing their MalOSS system~\cite{duan21}. This
dataset had some overlap with our basic corpus which we removed, leaving only
the unique data points for this experiment. Also, the dataset initially only
contained malicious packages; to balance it, we followed the same strategy as
for the basic corpus and added all benign versions of the contained packages. In
the end, this yielded a dataset with 372 package versions, out of which 40 were
malicious.%
\footnote{Detailed results for this experiment are included in the supplementary
materials.}

Based on the results from these two experiments, we will now answer the research
questions posed above.

\subsection{RQ1: Practical performance on newly published packages}
The results of the first experiment are presented in Table
~\ref{tab:sliding-window}. The table contains the date for which we collected
the package versions (\textbf{Date}) as well as the total number of versions
published on that date (\textbf{\# Versions}). Then, for each classifier, the
table contains the number of true positives (\textbf{\#TP}) and false positives
(\textbf{\#FP}) flagged by the classifier, annotated with the number of
additional true positives found by the clone detector ($+n$) and the number of
false positives eliminated by the reproducer ($-n$). As explained above, all true
positives were confirmed by the npm security team.

Thus, for example, the entry $16+2$ in the \textbf{\#TP} column for the decision
tree on August 4 means that the classifier flagged 16 true positive among the
17,063 package versions published that day of which the clone detector found two
additional copies. The entry $9-1$ in the \textbf{\#FP} column means that among
the nine false positives it flagged, one was successfully reproduced and hence
eliminated automatically.

As explained above, the clone detector is also treated like a fourth classifier. It
has no false positives and never misses identical copies, hence this column only
contains a single number per day. Note again that in this column we show the number of clones found on that same day, as opposed to the entries after the $+$ sign which depict the number of clones from the previous days. 

The first takeaway from this table is that the number of new packages published
every day is high, but quite variable, with almost four times as many packages
being published on July 29 (a Thursday) than on August 1 (a Sunday).

Secondly, we can see that all our classifiers are able to correctly classify
malicious package versions, with varying degrees of success. The decision tree
performs better than the rest, especially in terms of true positives. Removing
the overlap between classifiers, we were able to identify 95 previously unknown
malicious packages over the course of these seven days, which is a significant
number, especially considering that the entire set of malicious packages
detected prior to our work only contained 643 samples.

Third, we notice that on the first day all three classifiers produce an
unmanageable number of results. We therefore had to modify our approach and only
examined a subset of all flagged packages in detail, assuming all the rest to be
false positives. This means that the false-positive counts for this day are
likely to be overstated. However, once this set of packages is added to the
training set on the second day, the number of results drops dramatically, and by
the end of the week all three classifiers yield a relatively low number of
false positives.

Fourth, our results show that the reproducer has a low success rate in practice,
only being able to reproduce one or two packages on any given day. However,
given the overall low number of alerts towards the end of the week this is still
a valuable improvement. Similarly, clone detection only contributes a few
additional true positives each day (the 17 packages detected on August 4 being
an outlier, and mostly overlapping with the results from the decision tree), but
it still improves the overall results. Both mechanisms are computationally
inexpensive, and thus the help they provide comes at a low cost, making them
worth keeping in spite of their limited contributions. Conversely, this shows
that our classifiers add value beyond a purely textual scan looking for verbatim
copies of known malware.

In summary, we can answer RQ1 in the affirmative: \amalfi\ does indeed
detect malicious packages in practice. Further, since all the packages found by \amalfi\ and reported to npm had not been identified before, we can confidently claim that our approach complements existing solutions for malicious package detection in npm.

\newcommand{\pheader}{\textbf{Prec.}}
\newcommand{\rheader}{\textbf{Recall}}

\begin{table}[th]
\begin{tabular}{l||rr|rr|rr}
\textbf{Dataset} & \multicolumn{2}{c|}{\textbf{Decision Tree}} & \multicolumn{2}{c|}{\textbf{Naive Bayes}} & \multicolumn{2}{c}{\textbf{SVM}} \\
& \pheader & \rheader & \pheader & \rheader & \pheader & \rheader \\ \hline
Basic & 0.98 & 0.43 & 0.90 & 0.19 & (0.98) & (0.27) \\
MalOSS & 0.35 & 0.64 & 0.62 & 0.64 & 0.73 & 0.61 \\
\end{tabular}
\caption{Results from Experiment 2}
\label{tab:maloss-accuracy}
\end{table} 

\subsection{RQ2: Accuracy}
The results from Experiment 2 are presented in Table~\ref{tab:maloss-accuracy}.

The first row shows precision and recall measurements from the 10-fold
cross-validation experiment on our basic corpus, averaged over all ten runs. The
numbers for the SVM classifier have to be interpreted with care, since its $\nu$
parameter was fitted on this very dataset, hence we have put them in brackets.
All our models achieve very high precision, but the recall of Naive Bayes and
SVM is somewhat poor. This is expected due to the low prior of malicious packages. 

The second row shows precision and recall from running \amalfi\ on the
MalOSS dataset derived from the literature~\cite{duan21} as explained above. We
see that the recall is higher than with the previous row, at the expense of
precision. These results point to the trade-off between these two metrics, but
it is also worth pointing out that the MalOSS dataset contains a number of
packages labeled as malicious where npm disagreed with the authors' assessment
and did not take them down.

A more detailed comparison of \amalfi\ to MalOSS is unfortunately not
possible since they do not present statistics on false positives or performance.
The heavy-weight nature of their approach and its complicated setup involving a
sophisticated pipeline combining static and dynamic components made it
infeasible to run on our dataset.

Based on these results and the false-positive numbers discussed above, we can
give a cautiously positive answer to RQ2: \amalfi\ is reasonably precise and
does not produce an overwhelming number of results, making manual triaging of
results by a human auditor feasible. The second experiment suggests that there
may be a good number of false negatives, but at least the numbers for the
decision tree look promising.

\subsection{RQ3: Performance}
To characterize the performance of our approach, we measured three metrics: (i)
the time it takes to train the classifier, (ii) the time it takes to extract
features for a package version, and (iii) the time it takes to classify a
package version. All three measurements were obtained as a byproduct of Eperiment~1.

\begin{figure}[t!]
    \includegraphics[width=7cm]{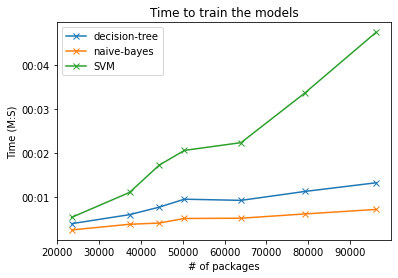}
\caption{Classifier training time}
\Description[Line graphs showing the training time for the three types of classifiers as the size of the training set increases.]{
The training time for all three classifiers is roughly linear in the size of the training set, with the decision tree and the
Naive Bayes classifier taking no more than a second to train even on the largest training set with over 90,000 samples. The SVM-based
classifier is slower to train, but still finishes in less than five seconds.
}
    \label{fig:training-time}
\end{figure}

Figure~\ref{fig:training-time} shows how long it took to train the classifiers
for each day of our experiment as a function of the size of the training set. As
can be seen, training is quite fast, taking no more than a few seconds despite
the steady increase in training data. The SVM-based classifier takes the longest
time, though it still seems to scale more or less linearly in the training-set
size. The other two classifiers are very quick to train, and it is clear that
the training set size could be increased substantially before training time
becomes a bottleneck.

To benchmark feature extraction, we post-processed logs from our run of
Experiment~1 to measure the time it takes to extract features for a randomly
chosen set of around 500 package versions. By and large, feature extraction takes less
than ten seconds, and for over half the packages considered it takes less than
one second. However, a single outlier package containing more than 11,000 files
takes more than ten minutes to extract, somewhat skewing the distribution for an
average extraction time of six seconds.

Lastly, we measured the time it took to predict whether a given package was
malicious. For all the classifiers, the time for prediction was less than a
second.

Based on these results, we can give a positive answer to RQ3: \amalfi\ is
fast enough for practical use.

\subsection{Threats to validity}
While the results of our evaluation are overall very promising, there are some
threats to the validity of our conclusions.

First, while the set of packages we considered in Experiment~1 was taken from
the wild, it may have been biased in ways that we did not anticipate, and so our
results may not generalize. Also, the basic corpus is to some degree biased in
that it contains clusters of similar malware samples resulting from copy-cat
campaigns.

Second, while we examined all packages flagged by \amalfi\ and reported the
true positives to npm, we limited ourselves to at most five minutes' inspection
time per package, which prevented detailed investigation of some of the larger
ones and may have caused us to miss true positives. Conversely, we may have been
mistaken in labelling some packages malicious, leading to missed false
positives, but this seems unlikely considering that npm have taken all reported
packages down, meaning that they agree with our assessment.

Finally, as noted above in our retraining step in Experiment~1 we assumed
packages that were not flagged by any classifier to be benign. This is not a
sound assumption in general, and might end up increasing the number of false
negatives over time. For this reason, and also to escape the slow but inexorable
rise in training time suggested by Figure~\ref{fig:training-time}, in practice
one would not want to continue retraining in this fashion indefinitely.
Table~\ref{tab:sliding-window} suggests diminishing returns from retraining
after a few days, but the data is clearly too sparse to draw a definite
conclusion.

\section{Discussion}\label{sec:discussion}
In this section we review the types of malicious packages our models found, take
a closer look at the models themselves, and finally touch upon tweaks to our
approach we investigated but were shown to be unsuccessful or unnecessary. 

\begin{figure}[ht]
\begin{minipage}[b]{0.45\linewidth}
\centering
\includegraphics[width=\textwidth]{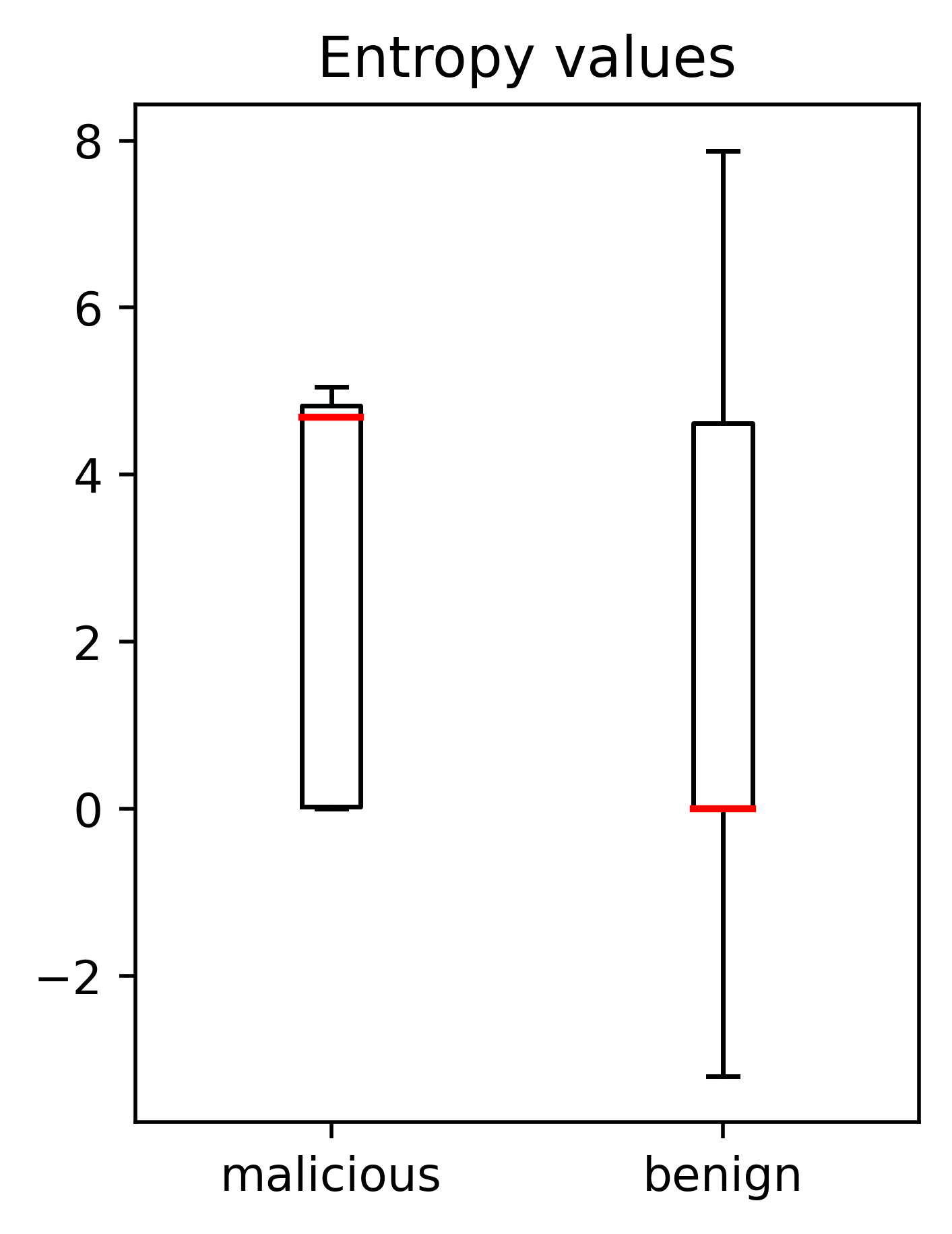}
\vspace{-0.4cm}
\label{fig:figure1}
\end{minipage}
\begin{minipage}[b]{0.5\linewidth}
\centering
\includegraphics[width=\textwidth]{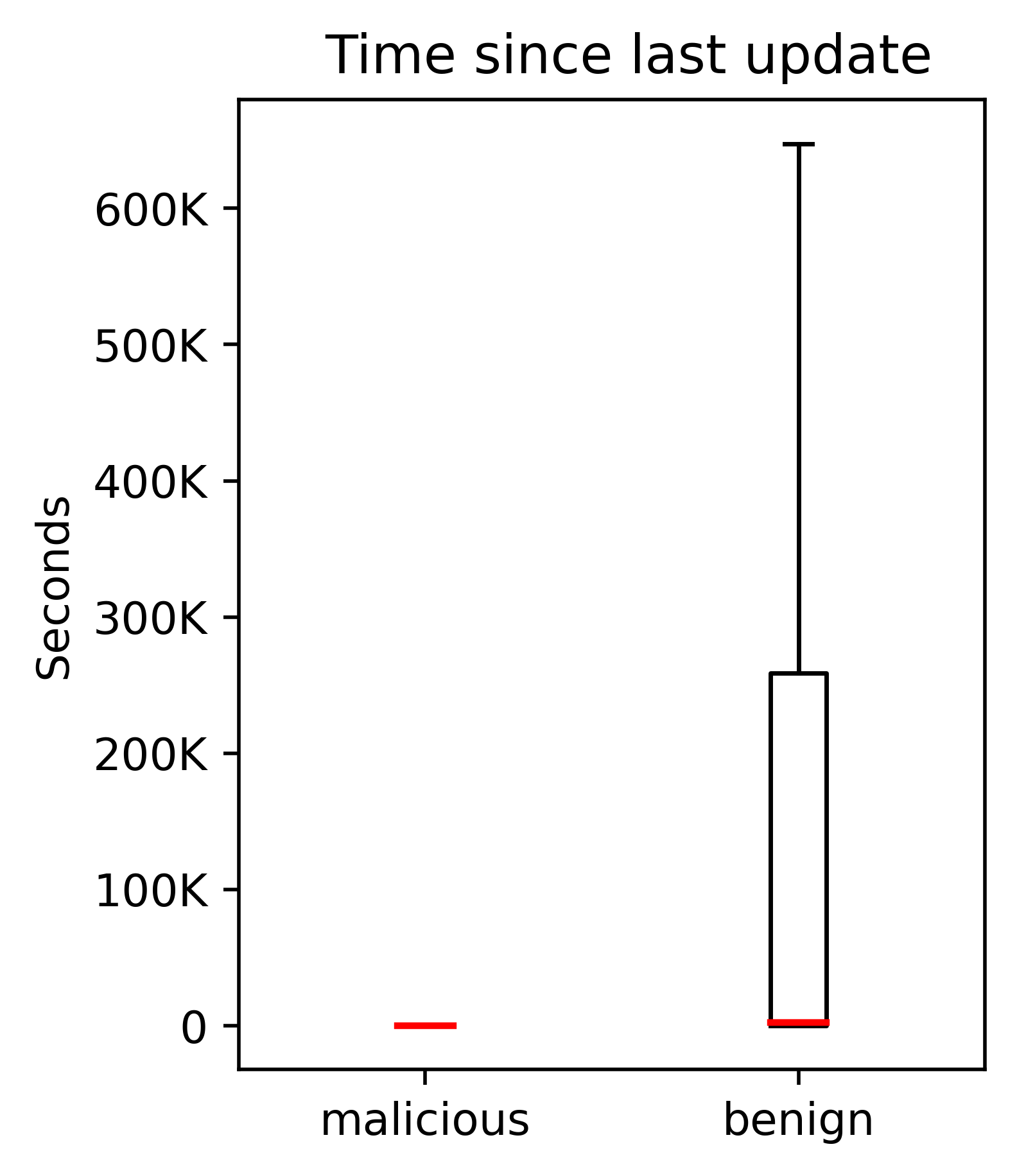}

\label{fig:figure2}
\end{minipage}
\caption{Entropy (left) and time (right) value distributions}
\Description[Distributions of (changes to) entropy values and time between updates for malicious and benign packages]{
    Median values for entropy are higher for malicious packages (>4) than for benign packages (barely more than zero), and less spread out.
    Median values for time between updates for malicious packages are very low, at just a few milliseconds.
}
\label{fig:entropy-time}
\end{figure}

Table~\ref{tab:discussion-features} details the types of discrete features
exhibited by the 95 malicious packages we found, while
Figure~\ref{fig:entropy-time} shows the distribution of the entropy and time
features of malicious and benign packages using boxplots.

\begin{table}[b]
\begin{tabular}{l|r}
\textbf{Feature} & \textbf{\# of packages} \\ \hline
File-system access & 11 \\
Process creation & 1 \\
Network access & 10 \\
Data encoding & 1 \\
Use of package installation scripts & 33 \\
Update type: major & 52 \\
Update type: minor & 1 \\
Update type: patch & 3 \\
Update type: prerelease & 9 \\
Update type: first & 30 \\
\end{tabular}
\caption{Features found in the 95 malicious packages}
\label{tab:discussion-features}
\end{table}

All packages used installation scripts or code in their main module to connect
to a remote host, with almost all of them sending PII to that host except for a
small handful of packages that only pinged the host without sending any
information, perhaps as a proof-of-concept or in preparation for an actual
attack. Our feature extractor did not detect the PII accesses themselves
(pointing to a need to improve our detection of this feature), but the packages
were detected anyway, usually because of the usage of the installation scripts
or network access. This suggests that our approach is robust enough to find
various types of malicious package versions. The fact that some features do not
appear in these specific packages does not necessarily indicate they are useless
or unnecessary; those features attempt to paint a general picture of
maliciousness in packages and they may prove to be useful in other batches.

A surprising observation in the data was the distribution of update types: the
majority of the malicious package versions we found were major updates,
contradicting our assumption that malicious package updates tend to ``hide''
behind a minor update. This could also mean that we missed malicious
package versions representing a minor update.

The distribution of average entropy values shows that the median, highlighted by
the thick \barcolor\ bar, is significantly higher for malicious packages
(4.69) than for benign packages (0.001), suggesting they are more
likely to contain minified code or binary files, in line with our expectation.
The time between updates also follows a rather expected distribution, with
malicious packages exhibiting a shorter median time between updates
(7.02 s) than the benign ones (2217.18 s).

\begin{figure}

\centering
\includegraphics[width=8cm]{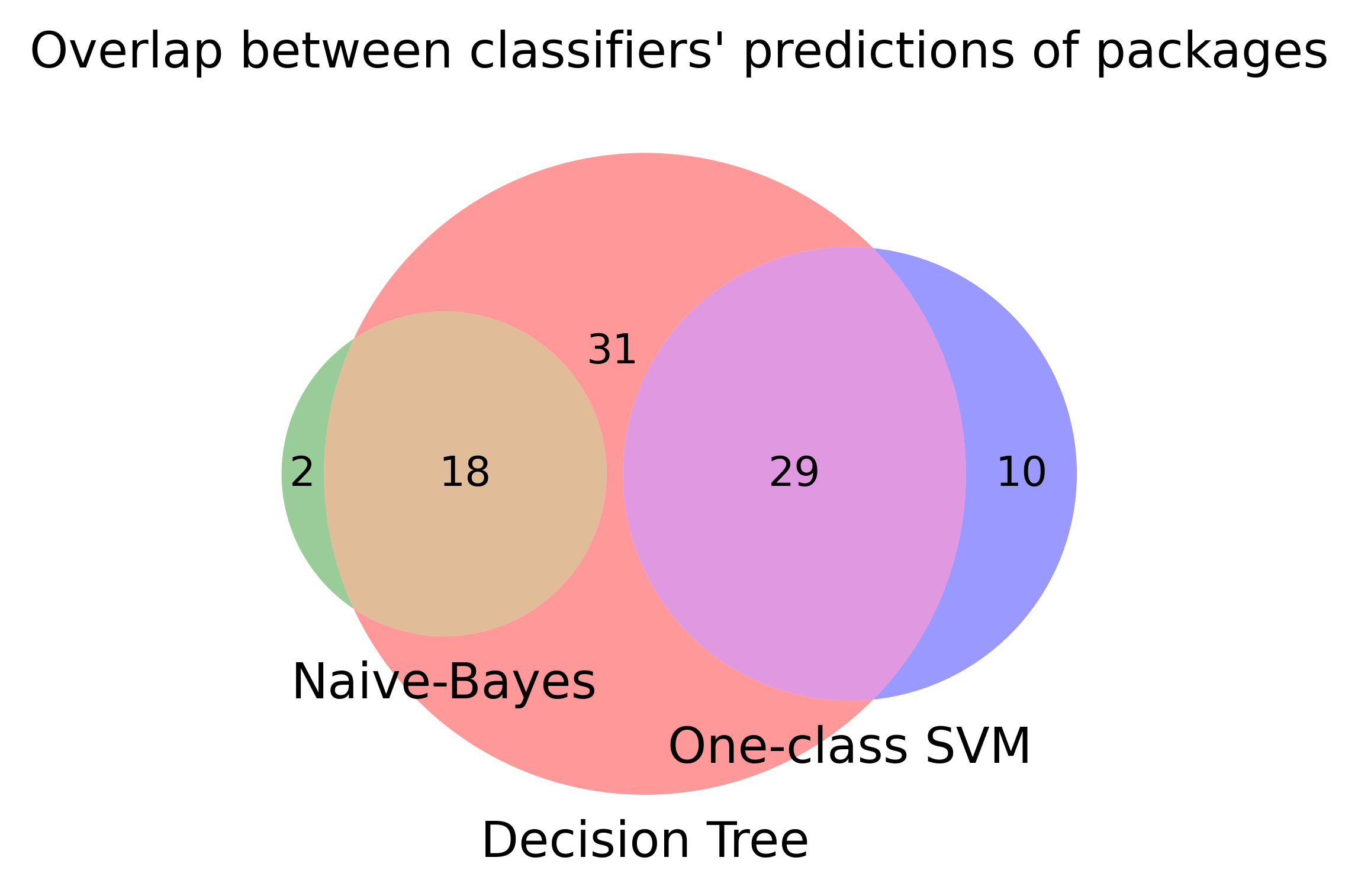}

\caption{Overlap among the three different classifiers on the 90 malicious packages they flag}
\Description[Diagram showing overlap between the three classifiers]{
    31 packages were only flagged by the decision tree, 10 only by SVM, and 2 only by Naive Bayes,
    while 29 packages were flagged by both decision tree and SVM, and 18 by both decision tree and Naive Bayes.
}
\label{fig:overlap}
\end{figure}

Next, we took a look at the generated classifiers and how their predictions
overlap. Figure~\ref{fig:overlap} shows a summary of the results on the 90
packages flagged by the classifiers (the remaining five having been flagged only
by the clone detector). While the decision tree takes the lion's share, each
individual algorithm makes its own contribution, suggesting that a combination
of all three might be a good choice in practice.

Since the decision tree classifiers are the ones that facilitate interpretation
we took a look at the features they use to make decisions. We noticed that the
classifiers for July 29 to July 31 examine all features except the one
representing uses of cryptographic functionality, and the remaining four
classifiers from August 1 onwards employ all eleven features, suggesting that
there is not much redundancy in our feature set.

Lastly, we tried out several tweaks that ultimately did not prove successful.
The literature often recommends using Random Forest classifiers instead of plain
decision trees, but we did not find them to provide any advantage in our
setting. We also investigated booleanizing features for the decision tree and
one-class SVM, but in both cases this led to worse performance, in the latter
case increasing the rate of false positives by more than 100\%.

\section{Related Work}\label{sec:related}
\begin{table*}[t]
\begin{tabular}{l|l|c|c}
\textbf{Category} & \textbf{Feature} & \textbf{\amalfi} & \textbf{Garrett et al.} \\ \hline
Access to PII & & \checkmark & \\ \hline
\multirow{3}{*}{Access to system resources} & File system access & \checkmark & \checkmark \\ \cline{2-4}
 & Process creation & \checkmark & \checkmark \\ \cline{2-4}
 & Network access & \checkmark & \checkmark \\ \hline
\multirow{3}{*}{Use of specific APIs} & Cryptographic functionality & \checkmark &  \\ \cline{2-4}
 & Data encoding & \checkmark &  \\ \cline{2-4}
 & Dynamic code generation & \checkmark & \checkmark \\ \hline
Use of installation scripts & & \checkmark & \checkmark \\ \hline
Presence of minified code and binary files & & \checkmark &  \\ \hline
Time between updates & & \checkmark &  \\ \hline
Update type & & \checkmark &  \\ \hline \hline
Added dependencies & &  & \checkmark \\ \hline
Added code & &  & \checkmark \\
\end{tabular}
\caption{Comparison of features considered in our models and those of Garrett et al.~\cite{garrett19}}
\label{tab:features-comparison}
\end{table*}

Our work has connections with four different research areas, which we survey
briefly: malicious package detection proper; malware and anomaly detection more
generally; package-registry security; and security implications of code reuse.

\paragraph*{Malicious-package detection}
Previous work in this area can be broadly divided into four categories:
general-purpose malicious-package detection approaches using machine
learning~\cite{garrett19} or program analysis~\cite{duan21,pfretzschner17,ohm20b};
techniques for rebuilding packages from source~\cite{goswami20,vu20,vu21}; and
finally work that specifically targets typosquatting~\cite{taylor20,vu20b}.

Garret et al.'s work on detecting malicious npm packages using machine-learning
techniques~\cite{garrett19} is very closely related to our work. They use a
k-means clustering algorithm to identify anomalous, and hence suspicious,
package updates.

Like us, they collect features for package updates, not just single package
versions, and the set of features they consider overlaps to some extent with
ours, as shown in Table~\ref{tab:features-comparison}. In particular, they also
consider access to system resources, dynamic code generation, and use of
installation scripts. We additionally consider access to PII and several
specific APIs as features, which they do not, though to some extent this is
covered by their feature recording added dependencies. Their feature set also
does not directly model the presence of minified code or binary files, though
again they do have a more general feature for added code that is similar in
spirit. They do not consider update type a feature, instead accounting for their
different characteristics by training separate models for each type of update.
Finally, they do not have any feature corresponding to our time between updates.

Our work is larger in scope than theirs, considering three different kinds of
classifiers instead of just one, and complementing the classifiers with package
reproduction and clone detection. Their evaluation on a set of 2288 package
updates suggests that their approach leads to many more alerts than ours,
flagging 539 updates as potentially suspicious; they did not triage the results
in detail, so it is unknown whether they succeeded in finding malicious
packages. Our experiments cover a much larger set of packages, and we have shown
that we can detect a significant number of previously unknown malicious
packages.

Duan et al.~\cite{duan21} study recent examples of supply-chain attacks,
pinpointing root causes and classifying attack vectors and malicious behaviors.
Based on their results, they built an analysis pipeline leveraging a combination of
static and dynamic program-analysis techniques to detect malicious packages
across three different package registries (npm,
PyPI,\footnote{\url{https://pypi.org/}} and
RubyGems\footnote{\url{https://rubygems.org/}}). In a large-scale experiment
covering more than one million packages, their approach identified 339
previously unknown malicious packages, 41 of them on npm. They do not provide
precise statistics on false positives or on the performance of their approach.

On the whole, our goals differ from theirs: where they aim to provide a
comparative framework for the security of registries, we specifically focus on
finding malicious npm packages. They update their detection rules manually as
results are assessed, while our classifiers can be retrained without further
manual effort beyond the assessment of results itself. Our approach seems
simpler and more lightweight than theirs, not requiring (potentially expensive)
deep static analysis or (potentially dangerous) code execution for dynamic
analysis.\footnote{While the reproducer does execute code, it only runs build
scripts, which are less likely to be malicious.} Nevertheless, we manage to find
more malicious packages on a smaller set than they do.

Pfretzschner et al.~\cite{pfretzschner17} propose the use of static analysis to
detect uses of JavaScript language features that can make a package vulnerable
to interference from a malicious downstream dependency. They present four
different attack scenarios involving global variables, monkey patching, and
caching of modules, though they did not find real-world examples of such
attacks.

Ohm et al.~\cite{ohm20b} propose a dynamic analysis for observing and measuring
the creation of artifacts during package installation as a way of detecting
malicious packages. While this is a promising research direction and potentially
practically very useful, approaches relying on code execution inherently tend to
be more heavyweight and harder to scale than our lightweight feature extraction.

At the shallower end of the analysis spectrum, tools like Microsoft Application
Inspector\footnote{\url{https://github.com/microsoft/ApplicationInspector}} and
OSSGadget\footnote{\url{https://github.com/microsoft/OSSGadget}} offer
regular-expression based scanning as a way of quickly detecting various types of
potential malware, including malicious packages. However, these tools tend to be
very noisy in practice and produce many false positives, precluding large-scale
usage.

Several researchers have proposed checking for differences between packages
hosted on registries and their purported source code as a way of detecting
malware. Goswami et al.~\cite{goswami20} report that this is difficult for npm
packages due to many irrelevant but non-malicious differences, an experience
that tallies with ours. Vu et al.~\cite{vu20,vu21} study the same problem for
PyPI, and similarly conclude that non-reproducibility by itself is a weak
indicator of maliciousness and needs to be combined with other techniques to
become effective, which is what we have done in this work.

For the specific problem of detecting typosquatting, Vu et al.~\cite{vu20b}
propose using edit distance as a metric for finding packages whose name is very
similar to another, while Taylor et al.~\cite{taylor20} employ a combination of
lexical similarity and package popularity. Our work does not specifically focus
on typosquatting, but may still be able to identify such packages from other
criteria. 

\paragraph*{Anomaly detection}
Malicious-package detection is a particular instance of the more general problem
of malware detection, which in turn is often phrased in terms of anomaly
detection~\cite{omar13}, where malware is characterized as anomalous outliers in
a larger set of benign samples. This framework has been brought to bear in a
wide variety of contexts, including detecting anomalous commits on
GitHub~\cite{goyal18,gonzalez21}, as well as malicious
websites~\cite{kazemian15}, binaries~\cite{vu19}, and mobile apps~\cite{cai20,
chen15, avdiienko15}.

Applying machine-learning techniques in these domains often faces the problem of
imbalanced datasets just like in our case. This line of research pointed us to
the advantages of using decision trees~\cite{omar13},
Naive-Bayes~\cite{kazemian15}, and One-class SVMs~\cite{omar13} as the base
algorithms for our models. More complicated models based on neural
networks~\cite{vu19} were not suitable for our problem given the relatively
small dataset in our possession. However, with more and better data, this could
be an avenue for future research. 

\paragraph*{Package-registry security}
The security mechanisms of the package registries and the impact of malicious
packages on these registries have also been studied extensively. Ohm et al.'s
study~\cite{ohm20} explores the forms attacks can have on different registries.
Others have focused on specific registries such as PyPI~\cite{ruohonen21,
bagmar21, alfadel21}, or npm~\cite{zimmermann19}.  On one hand, mechanisms to
understand the impact of malicious packages have been proposed~\cite{zerouali21,
nielsen21}. On the other hand, studies have also focused on ways registries can
mitigate the impact of malicious packages~\cite{ferreira21}. Compared to these
longer-term solutions (which are, of course, well worth investigating), our
focus is on short-term mitigation measures that do not require any changes to
the registries themselves, as explained in Section~\ref{sec:intro}.

\paragraph*{Code reuse and security}
Our work touches upon the risks of code reuse, which has recently seen a fair
amount of interest in the research literature.

Wang et al.~\cite{wang20} investigate third-party library usage in Java, with
particular focus on the problem of outdated libraries that may be lacking recent
bug fixes. They find that maintainers of client projects are often unwilling to
update to more recent library versions even when alerted to severe bugs in the
version they depend on. Prana et  al.~\cite{prana21} report similar results from
a study covering vulnerable dependencies in Java, Python, and Ruby.
Interestingly, they conclude that different levels of development activity,
project popularity, and developer experience do not affect the handling of
vulnerable-dependency reports.

Mirhoseini et al.~\cite{mirhoseini17} find that automated upgrade pull requests
improve the situation to some extent, although they can also have the adverse
side effect of overwhelming maintainers with upgrade notifications. For the case
of malicious npm packages, this is less of a problem since they are taken down
upon discovery and hence can no longer be depended on.

Gkortzis et al.~\cite{gkortzis21} specifically examine the relationship between
software reuse and security vulnerabilities. As one might expect, they find that
the larger a project the more likely it is to be affected by security
vulnerabilities, and similarly that projects with many dependencies are more
exposed to security risks. While their work focusses on vulnerable code as
opposed to malware, it stands to reason that similar correlations exist in the
latter case.

To mitigate this problem, Koishybayev et al.~\cite{koishybayev20} propose a
static analyzer called Mininode that eliminates unused code and dependencies
from Node.js applications, thereby reducing their attack surface.

\section{Conclusion}\label{sec:conclusion}
We have presented \amalfi, an approach to detecting malicious npm packages based on a
combination of a classifier trained on known samples of malicious and benign npm
packages, a reproducer for identifying packages that can be rebuilt from source,
and a clone detector for finding copies of known malicious packages. The
classifier works on a set of features extracted using a light-weight syntactic
analysis, including information about the capabilities the package makes use of
and how these change between versions.

We have presented an evaluation of our approach employing three different kinds
of classifiers: decision trees, Naive Bayesian classifiers, and SVMs. In our
experiments, all three techniques succeeded in detecting previously unknown
malicious packages, with the decision tree outperforming the other two, though
each classifier contributed unique results. While all three classifiers produce
false positives, their precision can be improved dramatically through continuous
retraining as past predictions are triaged. We have also shown that training,
feature extraction, and classification are very fast, suggesting that \amalfi\
is practically useful.

For future work, we are planning on investigating deeper feature extraction that
goes beyond the purely syntactic approach we have used so far, perhaps employing
light-weight static analysis. We would also like to experiment with more
advanced clone-detection approaches to identify similar but not textually
identical copies of malicious packages. Another area worth exploring would be
how to combine results from multiple classifiers, perhaps in the form of a
ranking of results that could aid in manual triaging. Finally, it would be very
interesting to apply our techniques to other ecosystems such as PyPI or
RubyGems, which also suffer from malicious packages in much the same way as npm.

\begin{acks}
  The authors would like to thank the npm team and the GitHub Trust and Safety team for providing us with access to the corpus of malicious packages,
  and for facilitating our experiments. We would also like to thank them, as well as our GitHub colleagues Bas Alberts and Henry Mercer, Tom Zimmermann
  and Patrice Godefroid of Microsoft Research, Laurie Williams, and the entire GitHub Next team for valuable feedback and advice during our work on this paper. Finally, we would like to thank Prof. Nenad Medvidovi\'c for his help and feedback. 
\end{acks}

\bibliographystyle{ACM-Reference-Format}
\balance
\bibliography{paper}

\end{document}